# DESIGN OF THE SNS NORMAL CONDUCTING LINAC RF CONTROL SYSTEM

Amy Regan, Sung-il Kwon, Tony S. Rohlev, Yi-Ming Wang,
LANL, Los Alamos, NM 87544, USA
Mark S. Prokop, David W. Thomson; Honeywell FM&T

*Abstract*

The Spallation Neutron Source (SNS) is being designed for operation in 2004. The SNS is a 1 GeV machine consisting of a combination normal-conducting and super-conducting linac as well as a ring and target area.

The linac front end is a 402.5 MHz RFQ being developed by Lawrence Berkeley Lab. The DTL (at 402.5 MHz) and the CCL (at 805 MHz) stages are being developed by Los Alamos National Laboratory. The expected output energy of the DTL is 87 MeV and that of the CCL is 185 MeV. The RF control system under development for the linac is based on the Low Energy Demonstration Accelerator's (LEDA) control system with some new features. This paper will discuss the new design approach and its benefits. Block diagrams and circuit specifics will be addressed. The normal conducting RF control system will be described in detail with references to the super-conducting control system where appropriate.

## 1 RF CONTROL FUNCTION OVERVIEW

The RF system for the SNS linac is well described in M. Lynch's paper in these proceedings [Ref. 1]. Specifically of interest to the RF Control System (RFCS) is the fact that one control system is required for each klystron. The RF control system must support operation of 402.5 MHz and 805 MHz normal conducting (NC) cavities, as well as 805 MHz superconducting (SRF) cavities. The intent of the RF Control system design is to provide a system which requires minimal hardware changes to support all three cavity types. For each cavity type, the governing specification is to provide cavity field control within ±0.5% amplitude and ±0.5° phase.

The functions required of the RFCS are: Cavity Field Control, Cavity Resonance Control, HPRF Protection, and Reference generation and distribution. Figure 1 shows a block diagram of the RFCS. We have selected a VXIbus architecture for the RFCS.

The present design combines cavity field control and resonance control into a single double-wide VXIbus module. The HPRF Protect function will be performed by another VXIbus module. Both are supported by a Clock Distribution Module. Physically this design differs from its predecessor (the Accelerator Production of Tritium RF control system) where Resonance Control and Field Control were individual modules and the HPRF Protect function required a VXI module plus multiple outboard chassis. Experience with LEDA has showed us we can reduce the number of channels supported by the HPRF Protect circuitry in such a way as to perform all required functions in a single VXI module only. We have also seen that combining the Field and Resonance Control functions into a single VXIbus module reduces the amount of backplane cross-communication and simplifies module-to-module interconnections.

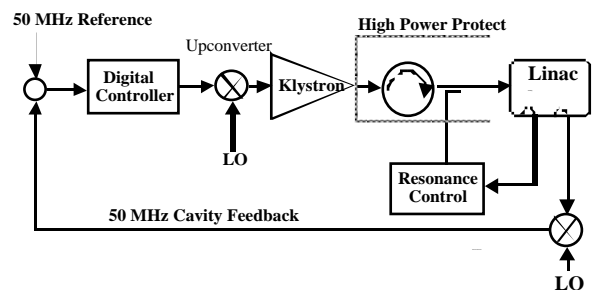

Figure 1: RF Controls for SNS.

A conceptual VXIbus crate layout is shown in Figure 2. Due to the physical separation of the klystrons for the NC cavities, we are only putting one RF Control system in a VXI crate. The SRF cavities' klystrons are located close enough together to encourage savings in crate and rack cost by co-locating two control systems in a single crate. We cannot do this for the NC systems because the distances between adjacent klystrons will detrimentally affect our control margin due to increased signal group delay.

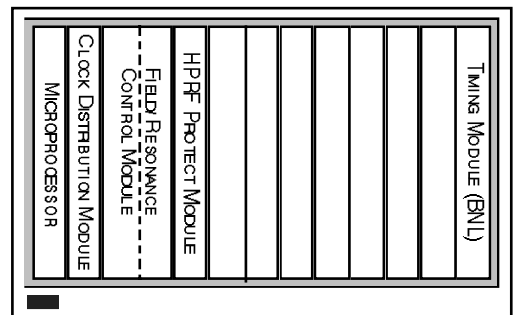

Figure 2. VXIbus RF Control System Crate Layout.

## 2 VXIBUS MODULE DESIGNS

### 2.1 Clock Distribution Module

The Clock Distribution Module (CDM) is quite similar to that of LEDA. It receives a phase stable reference signal from the Reference Distribution system and generates a 40 MHz ADC (analog-to-digital converter) clock (digital) and a 50 MHz IF (analog) for use by the Field/Resonance Control Module. A block diagram of the CDM is presented in Figure 3. The CDM also receives pulse timing information from the Brookhaven National Laboratory Timing Module and distributes it to the rest of the RFCS to synchronize the RFCS with the rest of the accelerator.

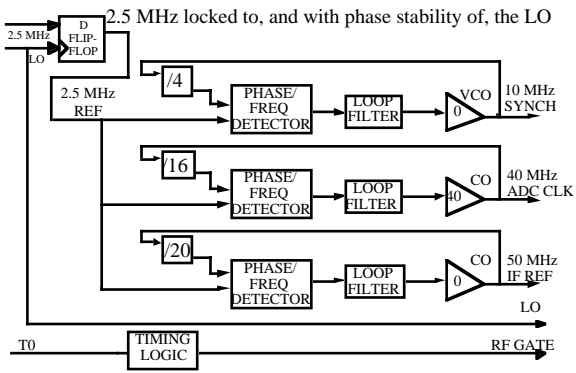

Figure 3. Clock Distribution Module Block Diagram.

### 2.2 Field/Resonance Control Module

The Field/Resonance Control Module has two primary functions. 1) It determines the current resonance condition of the cavity and sends a correction signal to the Cavity Resonance Control System (water cooling for the NC cavities) which brings the cavity back to resonance and maintains it. It also generates a frequency-shifted drive for conditions when the cavity is far off-resonance [Ref 2]). 2) The module also samples the cavity field and outputs the correct control signals for the klystron in order to keep the cavity field phase and amplitude within specification. It uses both PID (proportional-integral-derivative) control and an adaptive feedforward algorithm. The adaptive feedforward algorithm we refer to as an Iterative Learning Controller and is covered in a separate paper at these proceedings [Ref.3].

Figure 4 is a simplified block diagram of the Field/Resonance Control Module. This module makes extensive use of modern high-speed digital circuitry. Downconversion of the RF signals for the resonance control function and upconversion of the controlled RF drive signal for the klystron are the only analog circuits on the board. Significant digital components are two digital signal processors (DSPs) and four Complex Programmable Logic Devices (CPLDs). The two TI C60 family DSPs are used for the Resonance Control and Field Control functions (one each).

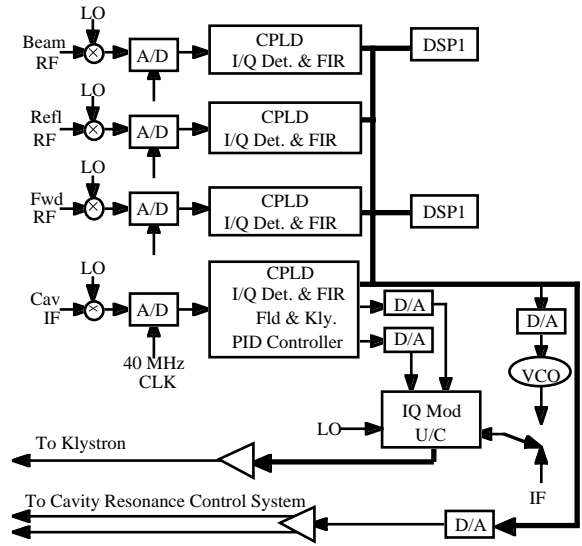

Figure 4. Simplified Control Module Block Diagram.

The Resonance Control algorithm is the same as used in LEDA [Ref. 2]. In the fast field control signal path, multi-rate digital processing is performed in CPLDs for optimized throughput. The Field Control DSP performs the slower, pulse-to-pulse, adaptive feedforward and gain scheduling features, while the Field Control CPLD does the actual fast feedback PID algorithm.

We will use the Altera EP20KE series CPLD family. Three of the four CPLDs are identical, containing a multiplexer/ multiplier (I/Q detector), digital filter, 2x2 rotation matrix and a PID controller). The I and Q output data rate is 20 MHz, and the expected delay/latency through the CPLD is 23.5 cycles (1174 ns). Figure 5 shows a block diagram of the CPLD signal flow and the associated delay/latency at each step. Note that the primary source of delay is the group delay through the FIR filter. The fourth CPLD in the Control Module includes this basic structure as well as the functionality to perform klystron phase control.

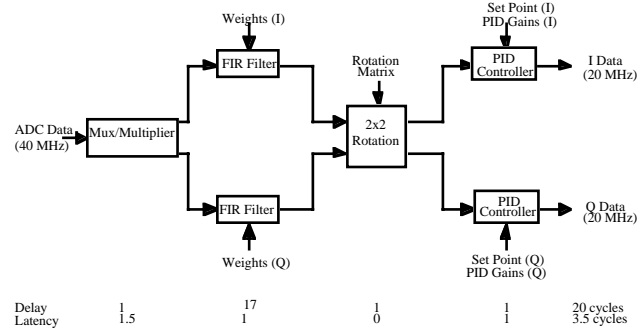

Figure 5. CPLD Signal Flow and Expected Delay.

## 2.3 HPRF Protect Module

The HPRF Protect function is fulfilled by a single VXIbus module. It is based on the HPRF Protect system for LEDA, which consists of a VXIbus module and four different chassis. Based on lessons learned on LEDA, we have been able to dramatically simplify this function. The module's purpose is to turn off the RF drive to the klystron should a fault occur within the HPRF transport line, be it arcing in the waveguide or unexpectedly high reflected power. There are six RF channels per module for monitoring RF power and ten inputs from the fiber optic arc subsystem for monitoring waveguide arcs. Instead of simply turning off the drive to the klystron on any given fault, logic is built into each channel to allow for a certain number of faults within a certain period of time (fault frequency) before declaring an RF-off state.

Besides waveguide arc monitoring *via* fiber optic detectors, logic is built into the module to interpret when a cavity arc occurs based on RF power signal from directional couplers at the cavity itself and the internal cavity field.

Like the Field/Resonance Control Module, the HPRF Protect Module is primarily digital. The analog front end consists of a 20 MHz bandwidth input filter at the RF frequency (402.5 or 805 MHz) and a true RMS power detector (AD8361). The input filter is the only part that is frequency-dependent. A six Ms/s analog-to-digital converter (Zilog XRD 6418 with 6 channels used) with 10 bits of resolution is used to digitize the power within the pulse. After that, all the comparators are digital, and with a simple Altera PLD for decode/actions, the expected total response time is 10 µs. A block diagram of a single RF channel (which is duplicated six times on the board) is given in Figure 6.

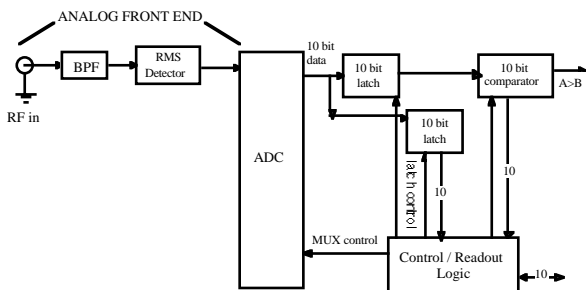

Figure 6. Single RF Channel Block Diagram.

## 3 REFERENCE DISTRIBUTION

The phase stable RF Reference required for the linac will be distributed via an insulated 3 1/8" coaxial line along the tunnel of the accelerator. The local oscillator frequency is distributed (352.5 MHz, or 755 MHz) and is mixed with the cavity field from individual pickup loops from each cavity in order to send a 50 MHz cavity field signal up to the klystron gallery where the RF Control electronics are located. The phase stability of the reference line is maintained to 0.1° at 755 MHz. [Ref. 4] A mockup of this system will be built at LANL later this year.

## 4 SUMMARY

The design of the RF Control System for the Normal Conducting SNS linac is well underway. Individual modules have been identified, specified, and are mostly through the initial design phase. In the next few months we will begin bread-boarding these modules and putting together an initial test system.